\documentclass{article}
\usepackage{mathrsfs}
\usepackage{amsfonts}
\usepackage{bbm}
\usepackage{amsmath, amsthm, amssymb}
\usepackage{color,calc}
\usepackage{hyperref}

\newcommand{\om}{\omega}

\begin{document}
 \vspace{15pt}

\begin{center}{\Large \bf Some new integrable systems constructed from the bi-Hamiltonian systems with pure differential Hamiltonian operators}
\end{center}
\begin{center}
{\it Yuqin Yao$^{1)}$\footnote{Corresponding author:
yqyao@math.tsinghua.edu.cn }, Yehui
Huang$^{2)}$\footnote{huangyh@mails.tsinghua.edu.cn}, Yuan
Wei$^{3)}$\footnote{wei-y79@126.com} and Yunbo
Zeng$^{2)}$\footnote{yzeng@math.tsinghua.edu.cn} }
\end{center}
\begin{center}{\small \it$^{1)}$Department of
  Applied Mathematics, China Agricultural University, Beijing, 100083, PR China\\
 $^{2)}$Department of Mathematical Science,
Tsinghua University, Beijing, 100084 , PR China\\
$^{3)}$Department of Mathematics and Information Science, Binzhou
University, Shandong 256603, PR China}
\end{center}

\vskip 12pt { \small\noindent\bf Abstract}
 {When both Hamiltonian operators of a bi-Hamiltonian system are pure differential operators, we show that the generalized Kupershmidt
 deformation (GKD) developed from the Kupershmidt deformation in \cite{kd} offers an useful way to construct new integrable system starting from
 the bi-Hamiltonian system. We construct some new integrable systems by means of the generalized Kupershmidt deformation
 in the cases of Harry Dym hierarchy, classical Boussinesq hierarchy and coupled KdV hierarchy. We show that the
 GKD of Harry Dym  equation,  GKD of classical Boussinesq equation and GKD of coupled KdV equation  are
 equivalent to the new integrable Rosochatius deformations of these soliton equations with self-consistent sources. We present the Lax Pair for these
 new systems. Therefore the generalized
Kupershmidt deformation provides a new way to construct new
integrable systems from bi-Hamiltonian systems and also offers a new
approach to obtain the Rosochatius deformation of soliton equation
with self-consistent sources. }
 \vskip 10pt
 { \small\noindent\bf Keywords:}{ Kupershmidt deformation; bi-Hamiltonian systems;
 Rosochatius deformation; soliton equation with self-consistent sources}
\vskip 12pt { \small\noindent\bf PACS: 02.30.Ik}

\section{Introduction}

In recent years the
 integrable deformations  of soliton hierarchies attracted a lot of
attention. Among them, the Rosochatius deformations of integrable
systems have important physical
applications\cite{Roso1}-\cite{Roso6}. Wojciechowski and Kubo et al.
researched the Rosochatius deformed Garnier system and Rosochatius
deformed  Jacobi  system, respectively\cite{Roso1,Roso2}. Zhou
generalize the Rosochatius method to study the integrable
Rosochatius deformations of some explicit constrained flows of
soliton equations \cite{Roso3,Roso4}. In  \cite{Roso5,Roso6}, we
proposed a systematic method to generalize the integrable
Rosochatius deformations of finite-dimensional integrable systems to
integrable Rosochatius deformations of soliton equations with
self-consistent sources.

On another hand it is known that one can construct a new integrable
system starting from a bi-Hamiltonian system. Fuchssteiner and Fokas
showed \cite{ff} that compatible symplectic structures lead in
natural way to hereditary symmetries, which provides a method to
construct a hierarchy of  exactly solvable evolution equations.
Olver and Rosenau \cite{OR} demonstrated that most integrable
bi-Hamiltonian systems are governed by a compatible trio of
Hamiltonian structures, and their recombination leads to integrable
hierarchies of nonlinear equations.

 For the following KdV6 equation or
nonholonomic deformation of KdV equation \cite{kdv6}
\begin{subequations}
\label{eqns:i1}
 \begin{align}
&u_{t}=u_{xxx}+6uu_{x}-\omega_{x},\\
& \omega_{xxx}+4u\omega_{x}+2u_{x}\omega=0,
 \end{align}
\end{subequations}
many authors studied it and established a lot of integrable
properties such as zero-curvature representation, bi-Hamiltonian
structure, conversed quantities, multisolitons and so
on\cite{kd}-\cite{yz0}. In particular, Kupershmidt  found \cite{kd}
that (\ref{eqns:i1}) can be converted into
\begin{subequations}
\label{eqns:i2}
 \begin{align}
 &u_{t}=J(\frac{\delta H_{3}}{\delta u})
 -J(\omega),\label{eqns:i2a}\\
& K(\omega)=0,\label{eqns:i2b}
 \end{align}
\end{subequations}
where
\begin{equation}
\label{eqns:i3}
J=\partial=\partial_{x},~K=\partial^{3}+2(u\partial+\partial u)
\end{equation}
are the two standard Hamiltonian operators of the KdV hierarchy
 and
$H_{3}=u^{3}-\frac{u_{x}^{2}}{2}.$ In general, for a bi-Hamiltonian
system
\begin{equation}
\label{eqns:i4} u_{t_{n}}=J(\frac{\delta H_{n+1}}{\delta
u})=K(\frac{\delta H_{n}}{\delta u}),
\end{equation}
the ansatz (\ref{eqns:i2}) provides a nonholonomic deformation of
bi-Hamiltonian systems\cite{kd}: $$u_{t_{n}}=J(\frac{\delta
H_{n+1}}{\delta u})
 -J(\omega),$$
\begin{equation} \label{eqns:5} K(\omega)=0
 \end{equation}
which is called as Kupershmidt deformation of bi-Hamiltonian
systems. This deformation is conjectured to preserve integrability
and the conjecture is verified in a few representative cases in
\cite{kd}. In \cite{nonho1}-\cite{nonho3}, the nonholonomic
deformations of mKdV equation, DNLS equation and KdV-type equations
were studied. In \cite{nonho4}, Zhou constructed the mixed soliton
hierarchy and show that the nonholonomic deformation of soliton
equations are some special numbers of the mixed soliton hierarchy.

 In \cite{yz0}, we showed that the
Kupershmidt deformation (\ref{eqns:i2}) of KdV equation is
equivalent to the Rosochatius deformation of KdV equation with
self-consistent sources, and presented the bi-Hamiltonian structure
for the Kupershmidt
 deformation (\ref{eqns:i2}). The conjecture is then proved in \cite{kkv} that
the Kupershmidt deformation of a bi-Hamiltonian system is itself
bi-Hamiltonian.
 Based on the Kupershmidt deformation
 (\ref{eqns:5}), we proposed the generalized Kupershmidt deformation
 for integrable bi-Hamiltonian systems in \cite{ayz}, which provides a new way to construct
 new integrable system from a bi-Hamiltonian system. Also we made
 the conjecture that the generalized Kupershmidt deformation
 for integrable bi-Hamiltonian systems preserves integrability.

In present paper, when both Hamiltonian operators of bi-Hamiltonian
system are pure differential operators, we show that
 the generalized Kupershmidt
 deformation (GKD) offers an useful way to construct new integrable systems starting from
 bi-Hamiltonian systems. By means of
 the generalized Kupershmidt deformation (GKD), we  construct some new  systems
  from some integrable bi-Hamiltonian systems in the cases of Harry Dym
 hierarchy, classical Boussinesq hierarchy and coupled KdV hierarchy. Then we show that
  these new  systems can be converted into the Rosochatius
  deformation of soliton equations with self-consistent sources,
   and their Lax pair can be found by using the method in \cite{Roso5,Roso6,cb2}.
  This indicates that the generalized Kupershmidt deformation  is equivalent to integrable Rosochatius
  deformation of soliton equations with self-consistent sources. Therefore
  the conjecture on integrability of the generalized Kupershmidt
  deformation is verified in these cases. This implies that the generalized Kupershmidt
  deformation  provides a new way to construct new
integrable systems from bi-Hamiltonian systems and also offers a new
 approach to construct the Rosochatius
  deformation of soliton equations with self-consistent sources in a
  different way from the method proposed in \cite{Roso5,Roso6}.
  However it remains  to study how to construct new integrable system
from the bi-Hamiltonian systems in which the Hamiltonian operators
are not pure differential operators, for example in the case of mKdV
hierarchy.
 In section 2, we construct  the generalized Kupershmidt
  deformation (GKD) of the Harry Dym hierarchy and  its Lax pair. We show that the GKD of Harry Dym equation
  is equivalent to the integrable Rosochatius
deformation of Harry Dym equation with self-consistent sources.
Section 3 is devoted to convert the GKD of classical Boussinesq
equation into the integrable Rosochatius deformation of classical
Boussinesq equation with self-consistent sources and present its Lax
pair.  Section 4 treats the GKD of coupled KdV equation, the new
coupled KdV equation  with self-consistent sources and it's Lax pair
are obtained. The last
  section presents the conclusion.

\section{The generalized Kupershmidt deformation of Harry Dym  hierarchy}
Consider a hierarchy of soliton equations with bi-Hamiltonian
structure
\begin{equation}
\label{eqns:0} u_{t_{n}}=J(\frac{\delta H_{n+1}}{\delta
u})=K(\frac{\delta H_{n}}{\delta u}),
\end{equation}
where $J$ and $K$ are two standard Hamiltonian operators. The
associated spectral problem reads
\begin{equation}
\label{eqns:00} L\phi=\lambda\phi.
\end{equation}
Assume that for $N$ distinct real eigenvalues $\lambda_{j}$, we have
$$L\varphi_{j}=\lambda_{j}\varphi_{j},~j=1,2,\cdots,N,$$
and
$$\frac{\delta \lambda_{j}}{\delta u}=f(\varphi_{j}).$$
Based on the  Kupershmidt deformation (5), we first generalize
Kupershmidt deformation  as follows
\begin{subequations}
\label{eqns:k2}
 \begin{align}
 &u_{t_{n}}=J(\frac{\delta H_{n+1}}{\delta u})
 -J(\sum_{j=1}^{N}\om_j),\label{eqns:k2a}\\
 & (\gamma_{j}J-\mu_{j}K)(\om_j)=0,~j=1,2,\cdots,
N,\label{eqns:k2b}
 \end{align}
\end{subequations}
where $\gamma_{j}$ and $\mu_{j}$ are constants, which also  gives to
 nonholonomic deformation of bi-Hamiltonian systems (\ref{eqns:0})
similar to the integrable KdV6's type of noholonomic deformation of
soliton equations. Furthermore, observe that $\omega_{j}$ in (8a) is
at the same position as $\frac{\delta H_{n+1}}{\delta u}$ and the
eigenvalues $\lambda_j$ are also the conserved quantity for
(\ref{eqns:0}) as $H_n$, it is reasonable to take
$\om_j=\frac{\delta \lambda_{j}}{\delta u}$ and this setting is
compatible with (8b). So we finally propose the generalized
Kupershmidt deformation for a bi-Hamiltonian systems in \cite{ayz}
as follows
\begin{subequations}
\label{eqns:k1}
 \begin{align}
 &u_{t_{n}}=J(\frac{\delta H_{n+1}}{\delta u}
 -\sum_{j=1}^{N}\frac{\delta
\lambda_{j}}{\delta u}),\label{eqns:k1a}\\
 & (\gamma_{j}J-\mu_{j}K)(\frac{\delta \lambda_{j}}{\delta
u})=0,~j=1,2,\cdots,N.\label{eqns:k1b}
 \end{align}
\end{subequations}
This deformation is also conjectured to preserve integrability and
the conjecture is verified in a few representative cases in
\cite{ayz}. In present paper, for some other specific cases, we will
further show that (\ref{eqns:k1}) gives rise to the Rosochatius
deformation of soliton equation with self-consistent sources or
soliton equation with self-consistent sources. Following the
procedure in [5,6,19], it is easy to find the Lax representation for
the Rosochatius deformation of soliton equation with self-consistent
sources, which implies the integrability of the Rosochatius
deformation of soliton equation with self-consistent sources or the
integrability of the generalized Kupershmidt deformation of soliton
equation.

We now consider the eigenvalue problem\cite{hd1}
\begin{equation}
\label{hd1} \left(\begin{array}{c}
\psi_{1}\\
\psi_{2}\\
 \end{array}\right)_{x}=U\left(\begin{array}{c}
\psi_{1}\\
\psi_{2}\\
 \end{array}\right),~~U=\left(\begin{array}{cc}
0 & 1\\
\lambda u & 0\\
 \end{array}\right).
\end{equation}
the associated  Harry Dym hierarchy reads\cite{hd2}
\begin{equation}
\label{hd3} u_{t_{n}}=J(\frac{\delta H_{n+1}}{\delta
u})=K(\frac{\delta H_{n}}{\delta u}),
\end{equation}
where $J=\partial^{3}$ and $K=u\partial+\partial u$ are two standard
Hamiltonian operators,~$H_{0}=-\int u dx,~H_{-1}=\int
2u^{\frac{1}{2}} dx,~H_{-2}=\int
\frac{1}{8}u^{-\frac{5}{2}}u_{x}^{2} dx,$ $H_{-3}=\int
\frac{1}{16}(\frac{35}{16}u^{-\frac{11}{2}}u_{x}^{4}-u^{-\frac{7}{2}}u_{xx}^{2}
)dx,\cdots.$

When $n=-2$, (\ref{hd3}) gives the Harry Dym equation
\begin{equation}
\label{hde} u_{t}=(u^{-\frac{1}{2}})_{xxx}.
\end{equation}

 Assume that for
$N$ distinct real $\lambda_{j}$, we have
\begin{equation}
\label{hd4} \left(\begin{array}{c}
\psi_{1j}\\
\psi_{2j}\\
 \end{array}\right)_{x}=\left(\begin{array}{cc}
0 & 1\\
\lambda_{j} u & 0\\
 \end{array}\right)\left(\begin{array}{c}
\psi_{1j}\\
\psi_{2j}\\
 \end{array}\right).\end{equation}
Its easy to find that
\begin{equation}
\label{lamda}\frac{\delta \lambda_{j}}{\delta
u}=-\lambda_{j}\psi_{1j}^{2}.\end{equation}

 Take $\gamma_{j}=1,~\mu_{j}=\frac{1}{2}\lambda_{j},~(j=1,\cdots,N)$ and $n=-2$, the generalized Kupershmidt deformation
 (\ref{eqns:k2})
 gives rise to the following new generalized Harry Dym equation
 \begin{subequations}
\label{eqns:i10}
 \begin{align}
&u_{t}=(u^{-\frac{1}{2}})_{xxx}-\sum_{j=1}^{N}\omega_{jxxx},\\&
\omega_{jxxx}-\lambda_{j}u\omega_{jx}-\frac{1}{2}\lambda_{j}u_{x}\omega_j=0,~j=1,2,\cdots,N.
 \end{align}
\end{subequations}
By replacing $\omega_j$ by (\ref{lamda}), (15b) yields
$$\psi_{1j}(\psi_{1jxx}-\lambda_{j}u\psi_{1j})_{x}+3\psi_{1jx}(\psi_{1jxx}-\lambda_{j}u\psi_{1j})=0,$$
which immediately leads to
$$\psi_{1jxx}-\lambda_{j}u\psi_{1j}=\frac{\mu_{j}}{\psi_{1j}^{3}},$$
where $\mu_{j},~j=1,2,\cdots,N$ are integral constants.
 Therefore (\ref{eqns:k1}) gives
 rise to the  following new generalized Kupershmidt deformation of Harry Dym
 equation(GKDHDE)
\begin{subequations}
\label{eqns:k20}
 \begin{align}
&u_{t}=(u^{-\frac{1}{2}})_{xxx}+\sum\limits_{j=1}^{N}\lambda_{j}(\psi_{1j}^{2})_{xxx},\\
&
\psi_{1jxx}+(u-\lambda_{j})\psi_{1j}=\frac{\mu_{j}}{\psi_{1j}^{3}},
~j=1,2,\cdots,N
 \end{align}
\end{subequations}
which can  be regarded as the Rosochatius deformation of Harry Dym
equation with self-consistent sources (RD-HDESCS). When
$\mu_{j}=0,~j=1,\cdots,N,$ (\ref{eqns:k20}) reduces to the Harry Dym
equation with self-consistent sources(HDESCS).  In the following, we
derive the Lax pair for (\ref{eqns:k20}). Fist we derive the Lax
pair for HDESCS.

Setting $\psi_{1}=\psi, ~\psi_{2x}=\lambda u\psi_{1}$ and comparing
to the Harry Dym equation, we can assume the Lax representation of
the HDESCS has the form
\begin{subequations}
\label{eqns:spectralproblem2}
 \begin{align}
&\psi_{xx}=\lambda u\psi, \label{eqns:spectralproblem2a}\\
& \psi_{t}=-\frac{1}{2}B_{x}\psi+B\psi_{x},\label{eqns:spectralproblem2b}\\
&B=-2u^{-\frac{1}{2}}\lambda+\sum\limits_{j=1}^{N}\frac{\alpha_{j}f(\psi_{j})}{\lambda-\lambda_{j}}+
\sum\limits_{j=1}^{N}\beta_{j}f(\psi_{j}),
\label{eqns:spectralproblem2c}
 \end{align}
\end{subequations}
where  $f(\psi_{j})$ is undetermined function of $\psi_{j}$.
 The compatibility
condition of (\ref{eqns:spectralproblem2a}) and
(\ref{eqns:spectralproblem2b}) gives
\begin{equation}
\label{eqns:condition}
       u_{t}\lambda=LB+(2B_{x}u+Bu_{x})\lambda,
\end{equation}
where $L=-\frac{1}{2}\partial^{3}$. Then
(\ref{eqns:spectralproblem2c}) and (\ref{eqns:condition}) yields
$$u_{t}\lambda=(u^{-\frac{1}{2}})_{xxx}\lambda+\sum\limits_{j=1}^{N}\frac{\alpha_{j}}
{\lambda-\lambda_{j}}[-\frac{1}{2}f^{'''}\psi_{jx}^{3}+(-\frac{3}{2}\lambda_{j}uf^{''}\psi_{j}+\frac{3}{2}\lambda_{j}uf')\psi_{jx}$$
$$+\lambda_{j}u_{x}(f-\frac{1}{2}\psi_{j}f')]
+\sum\limits_{j=1}^{N}\beta_{j}[-\frac{3}{2}u\psi_{j}\psi_{jx}f^{''}-\frac{1}{2}u_{x}\psi_{j}f^{'}-\frac{1}{2}u
\psi_{jx}f^{'}+2u\psi_{jx}f^{'}+u_{x}f]\lambda$$
\begin{equation}
\label{eqns:expand}
+\sum\limits_{j=1}^{N}[\beta_{j}\lambda_{j}(-\frac{3}{2}\psi_{j}\psi_{jx}f^{''}-\frac{1}{2}u_{x}\psi_{j}f^{'}-\frac{1}{2}u\psi_{jx}f^{'})+2u\lambda_{j}\alpha_{j}\psi_{jx}f^{'}+\lambda_{j}\alpha_{j}u_{x}f]~~~~~~~~~
~~~~~~~~~~~~~~~~~~~~~
\end{equation}
Here   $f'$ denotes the partial derivative of the function $f$ with
respect to the variable $\psi_{j}$. In order to determine
$f,~\alpha_{j}$ and $\beta_{j}$, we compare the coefficients of
$\frac{1} {\lambda-\lambda_{j}},~\lambda$ and $\lambda^{0}$,
respectively. We first observe the coefficients of $\frac{1}
{\lambda-\lambda_{j}}$. The coefficients of $\psi_{jx}^{3}$ gives
\begin{equation}
\label{eqns:f1} f^{'''}=0,
\end{equation}
 The coefficients of $\psi_{jx}$ gives
\begin{equation}
\label{eqns:f2} f^{''}\psi_{j}-f'=0,
\end{equation}
The other terms gives
\begin{equation}
\label{eqns:f3} \frac{1}{2}f'\psi_{j}-f=0.
\end{equation}
From (\ref{eqns:f1}), (\ref{eqns:f2}) and (\ref{eqns:f3}) we obtain
$f=\psi_{j}^{2}$. Substituting $f=\psi_{j}^{2}$ into the
coefficients of $\lambda$ gives
$$u_{t}=(u^{-\frac{1}{2}})_{xxx}+\sum\limits_{j=1}^{N}\beta_{j}(4u\psi_{j}\psi_{jx}+u_{x}\psi_{j}^{2}).$$
Comparing with the HDESCS we can determine
$$\beta_{j}=-2\lambda_{j}^{2}.$$
 Substituting $f=\psi_{j}^{2}$ and $\beta_{j}=-2\lambda_{j}^{2}$ into the
coefficients of $\lambda^{0}$ gives
$$\sum\limits_{j=1}^{N}[(2\lambda_{j}^{3}+\lambda_{j}\alpha_{j})(4u\psi_{j}\psi_{jx}+u_{x}\psi_{j}^{2})]=0$$
which gives
$$\alpha_{j}=-2\lambda_{j}^{2}.$$
Thus we obtained the Lax pair of the HDESCS
\begin{subequations}
\label{eqns:twolaxp}
 \begin{align}
&\psi_{xx}=\lambda u\psi,\\ \nonumber &
\psi_{t}=(-\frac{1}{2}u^{-\frac{3}{2}}u_{x}\lambda-2\sum\limits_{j=1}^{N}\lambda_{j}^{2}\psi_{j}\psi_{jx}
-2\sum\limits_{j=1}^{N}\frac{\lambda_{j}^{2}\psi_{j}\psi_{jx}}{\lambda-\lambda_{j}})\psi\\
&~~~~~~~+(-2u^{-\frac{1}{2}}\lambda+2\sum\limits_{j=1}^{N}\lambda_{j}^{2}\psi_{j}^{2}
+2\sum\limits_{j=1}^{N}\frac{\lambda_{j}^{2}\psi_{j}^{2}}{\lambda-\lambda_{j}})\psi_{x},
 \end{align}
\end{subequations}
which equivalent to (\ref{hd1}) and
\begin{equation}
\label{eqns:v} \left(\begin{array}{c}
\psi_{1}\\
\psi_{2}\\
 \end{array}\right)_{t}=V\left(\begin{array}{c}
\psi_{1}\\
\psi_{2}\\
 \end{array}\right),~\psi_{1}=\psi,~\psi_{2}=\psi_{1x}\end{equation}
with $$V=\left(\begin{array}{cc}
-\frac{1}{2}u_{x}u^{-\frac{3}{2}}\lambda & -2u^{-\frac{1}{2}}\lambda\\
-2u^{\frac{1}{2}}\lambda^{2}-\frac{1}{2}(u_{xx}u-\frac{3}{2}u_{x}^{2})u^{-\frac{5}{2}}\lambda & \frac{1}{2}u_{x}u^{-\frac{3}{2}}\lambda\\
 \end{array}\right)+2\sum\limits_{j=1}^{N}\left(\begin{array}{cc}
0 & 0\\
u\psi_{1j}^{2}\lambda_{j}^{2}\lambda  & 0\\
 \end{array}\right)$$$$-2\sum\limits_{j=1}^{N}\frac{\lambda_{j}^{2}\lambda}{\lambda-\lambda_{j}}\left(\begin{array}{cc}
\psi_{1j}\psi_{2j} & -\psi_{1j}^{2}\\
\psi_{2j}^{2}  & -\psi_{1j}\psi_{2j}\\
 \end{array}\right).~~~~~~~~~~~~~~~~~~~~~~~~~~~~~~~~~~~~~~$$

 Finally, as proposed in \cite{Roso5,Roso6}, the above formula induces the Lax
 pair (10) and (24) for RD-HDESCS (\ref{eqns:k20}) with
 $$V=\left(\begin{array}{cc}
-\frac{1}{2}u_{x}u^{-\frac{3}{2}}\lambda & -2u^{-\frac{1}{2}}\lambda\\
-2u^{\frac{1}{2}}\lambda^{2}-\frac{1}{2}(u_{xx}u-\frac{3}{2}u_{x}^{2})u^{-\frac{5}{2}}\lambda & \frac{1}{2}u_{x}u^{-\frac{3}{2}}\lambda\\
 \end{array}\right)+2\sum\limits_{j=1}^{N}\left(\begin{array}{cc}
0 & 0\\
u\psi_{1j}^{2}\lambda_{j}^{2}\lambda  & 0\\
 \end{array}\right)$$$$-2\sum\limits_{j=1}^{N}\frac{\lambda_{j}^{2}\lambda}{\lambda-\lambda_{j}}\left(\begin{array}{cc}
\psi_{1j}\psi_{2j} & -\psi_{1j}^{2}\\
\psi_{2j}^{2} +\frac {\mu_j}{\psi_{1j}^2} & -\psi_{1j}\psi_{2j}\\
 \end{array}\right).~~~~~~~~~~~~~~~~~~~~~~~~~~~~~~~~~~~~~~$$

\section{The generalized Kupershmidt deformation of the classical Boussinesq hierarchy}
For the classical Boussinesq eigenvalue problem \cite{cb1}
\begin{equation}
\label{j1} \left(\begin{array}{c}
\psi_{1}\\
\psi_{2}\\
 \end{array}\right)_{x}=U\left(\begin{array}{c}
\psi_{1}\\
\psi_{2}\\
 \end{array}\right),~~U=\left(\begin{array}{cc}
0 & 1\\
-\lambda^{2}+\lambda v+u-\frac{1}{4}v^{2} & 0\\
 \end{array}\right),
\end{equation}
the associated classical Boussinesq hierarchy reads
$$\left(\begin{array}{c}
u\\
v\\
 \end{array}\right)_{t_{n}}=J\left(\begin{array}{c}
b_{n+1}\\
b_{n+2}-\frac{1}{2}vb_{n+1}\\
 \end{array}\right)=J\left(\begin{array}{c}
\frac{\delta H_{n+1}}{\delta
u}\\
\frac{\delta H_{n+1}}{\delta
v}\\
 \end{array}\right)=K\left(\begin{array}{c}
\frac{\delta H_{n}}{\delta
u}\\
\frac{\delta H_{n}}{\delta
 v}\\
 \end{array}\right)$$
where $$J=\left(\begin{array}{cc}
0 & 2\partial\\
2\partial &0\\
 \end{array}\right),~K=\left(\begin{array}{cc} -\frac{1}{2}\partial^{3}+u\partial+\partial u & v\partial\\
\partial v & 2\partial\\
 \end{array}\right),$$$$\left(\begin{array}{c}
b_{n+2}\\
b_{n+1}\\
 \end{array}\right)=L\left(\begin{array}{c}
b_{n+1}\\
b_{n}\\
 \end{array}\right),
 $$ $$L=\left(\begin{array}{cc}
v-\frac{1}{2}\partial^{-1}v_{x} & -\frac{1}{4}\partial^{2}-\frac{1}{2}\partial^{-1}(u-\frac{1}{4}v^{2})_{x}+u-\frac{1}{4}v^{2}\\
1 &0\\
 \end{array}\right),$$
 $$b_0=b_1=0,~b_2=2,~n=1,2,\cdots.$$
 For $N$ distinct real $\lambda_{j}$, from the spectral problem
$$\psi_{1jx}=\psi_{2j},~\psi_{2jx}=(-\lambda_{j}^{2}+\lambda_{j} v+u-\frac{1}{4}v^{2})\psi_{1j}$$
we have
\begin{equation}
\label{lamdab}\frac{\delta \lambda_{j}}{\delta u}=-\psi_{1j}^{2},~
\frac{\delta \lambda_{j}}{\delta
v}=(-\lambda_{j}+\frac{1}{2}v)\psi_{1j}^{2}.\end{equation}
 When take $\gamma_{j}=-\lambda_{j},~\mu_{j}=-1,$
 and $n=4$,  the generalized Kupershmidt deformation
 (\ref{eqns:k2}) gives rise to the following new generalized classical Boussinesq
equation
\begin{subequations}\label{aj} \begin{align}\nonumber
&u_{t}=-\frac{3}{4}vv_{xxx}-\frac{1}{2}u_{xxx}+\frac{3}{2}v^{2}u_{x}+3uvv_{x}+3uu_{x}-\frac{3}{2}v_{x}v_{xx}\\
&~~~~~~~~-2\sum_{j=1}^{N}\omega_{1jx},\\
&v_{t}=-\frac{1}{2}v_{xxx}+3vu_{x}+\frac{3}{2}v^{2}v_{x}+3uv_{x}-2\sum_{j=1}^{N}\omega_{2jx},\\
&-\frac{1}{2}\omega_{1jxxx}+2u\omega_{1jx}+u_{x}\omega_{1j}+v\omega_{2jx}-2\lambda_{j}\omega_{2jx}=0,\\
&v_{x}\omega_{1j}+v\omega_{1jx}+2\omega_{2jx}-2\lambda_{j}\omega_{1jx}=0,
 ~j=1,2,\cdots,N.
\end{align}
\end{subequations}
The generalized Kupershmidt deformation of the classical Boussinesq
hierarchy is constructed from  (\ref{eqns:k1}) as follows
\begin{subequations}
\label{eqns:j2}
 \begin{align}
 &\left(\begin{array}{c}
u\\
v\\
 \end{array}\right)_{t_{n}}=J(\left(\begin{array}{c}
\frac{\delta H_{n+1}}{\delta
u}\\
\frac{\delta H_{n+1}}{\delta
 v}\\
 \end{array}\right)
 -\sum_{j=1}^{N}\left(\begin{array}{c}
\frac{\delta \lambda_{j}}{\delta
u}\\
\frac{\delta \lambda_{j}}{\delta
v}\\
 \end{array}\right))
 ,\label{eqns:j2a}\\
& (K-\lambda_{j}J)\left(\begin{array}{c} \frac{\delta
\lambda_{j}}{\delta
u}\\
\frac{\delta \lambda_{j}}{\delta
v}\\
 \end{array}\right)=0,~j=1,2,\cdots,N.\label{eqns:j2b}
 \end{align}
\end{subequations}
By substituting (\ref{lamdab}), (\ref{eqns:j2b}) yield
$$\psi_{1j}(\psi_{2jxx}-u\psi_{1j}-\lambda_{j}v\psi_{1j}+\frac{1}{4}v^{2}\psi_{1j}+\lambda_{j}^{2}\psi_{1j})_x$$
$$
+3\psi_{1jx}(\psi_{2jxx}-u\psi_{1j}-\lambda_{j}v\psi_{1j}+\frac{1}{4}v^{2}\psi_{1j}+\lambda_{j}^{2}\psi_{1j})=0,$$
which leads to
$$\psi_{2jxx}=(-\lambda_{j}^{2}+\lambda_{j}
v+u-\frac{1}{4}v^{2})\psi_{1j}+\frac{\mu_{j}}{\psi_{1j}^{3}},
~j=1,2,\cdots,N.$$ Take $n=4$, the generalized Kupershmidt
deformation of classical Boussinesq equation (\ref{eqns:j2}) gives
rise to the following new system

\begin{subequations}\label{j3} \begin{align}\nonumber
&u_{t}=-\frac{3}{4}vv_{xxx}-\frac{1}{2}u_{xxx}+\frac{3}{2}v^{2}u_{x}+3uvv_{x}+3uu_{x}-\frac{3}{2}v_{x}v_{xx}\\
&~~~~~~~~-\sum_{j=1}^{N}(-4\lambda_{j}\psi_{1j}\psi_{1jx}+v_{x}\psi_{1j}^{2}+2v\psi_{1j}\psi_{1jx}),\\
&v_{t}=-\frac{1}{2}v_{xxx}+3vu_{x}+\frac{3}{2}v^{2}v_{x}+3uv_{x}+4\sum_{j=1}^{N}\psi_{1j}\psi_{1jx},\\
&\psi_{2jxx}=(-\lambda_{j}^{2}+\lambda_{j}
v+u-\frac{1}{4}v^{2})\psi_{1j}+\frac{\mu_{j}}{\psi_{1j}^{3}},
~j=1,2,\cdots,N
\end{align}
\end{subequations}
which  is  regarded as the Rosochatius deformation of classical
Boussinesq equation with self-consistent sources. Following the
procedure in \cite{Roso5,Roso6,cb2}, we can find that Eq.(\ref{j3})
has the Lax representation (\ref{hd1}) and (\ref{eqns:v}) with
$$U=\left(\begin{array}{cc}
0 & 1\\
-\lambda^{2}+\lambda v+u-\frac{1}{4}v^{2} & 0\\
 \end{array}\right),$$
$$V=\left(\begin{array}{cc}
 -\frac{1}{2}v_{x}\lambda -\frac{1}{2}vv_{x}-\frac{1}{2}u_{x}& 2\lambda^{2}+v\lambda+\frac{1}{2}v^{2}+u\\
F
 &\frac{1}{2}v_{x}\lambda +\frac{1}{2}vv_{x}+\frac{1}{2}u_{x}\\
 \end{array}\right)
 $$
 $$+\sum\limits_{j=1}^{N}\left(\begin{array}{cc}
 0& 0\\
(\lambda_{j}-v+\lambda) \psi_{1j}^{2} &0\\
 \end{array}\right)-\sum\limits_{j=1}^{N}\frac{1}{\lambda-\lambda_{j}}
 \left(\begin{array}{cc}
 \psi_{1j} \psi_{2j}& - \psi_{1j}^{2}\\
\psi_{2j}^{2}+\frac{\mu_{j}}{\psi_{1j}^{2}}&-\psi_{1j} \psi_{2j}\\
 \end{array}\right)$$
where
$$F=-2\lambda^{4}+v\lambda^{3}+u\lambda^{2}+(-\frac{1}{2}v_{xx}+\frac{1}{4}v^{3}+2uv)\lambda-\frac{1}{2}v_{x}^{2}
-\frac{1}{2}vv_{xx}-\frac{1}{2}u_{xx}+\frac{1}{4}uv^{2}+u^{2}-\frac{1}{8}v^{4}.$$

\section{The generalized Kupershmidt deformation of the coupled KdV hierarchy}
The  coupled KdV hierarchy (cKdVH) and its Backlund transformations
was derived by Levi\cite{ckdv1}. This hierarchy has two important
features: firstly the odd sub-hierarchy can be reduced to ordinary
KdV hierarchy and secondly its third member resembles  the
celebrated Hirota-Satsuma system of equations\cite{ckdv2,ckdv3}.

The coupled KdV  eigenvalue problem reads\cite{ckdv4,ckdv5}
\begin{equation}
\label{ckv1} \left(\begin{array}{c}
\psi_{1}\\
\psi_{2}\\
 \end{array}\right)_{x}=U\left(\begin{array}{c}
\psi_{1}\\
\psi_{2}\\
 \end{array}\right),~~U=\left(\begin{array}{cc}
-\frac{1}{2}(\lambda-u) & -v\\
1 & \frac{1}{2}(\lambda-u)\\
 \end{array}\right),
\end{equation}
the associated cKdVH can be written as the bi-Hamiltonian structure
$$\left(\begin{array}{c}
u\\
v\\
 \end{array}\right)_{t_{n}}=J\left(\begin{array}{c}
a_{n+1}\\
-c_{n+1}\\
 \end{array}\right)=J\left(\begin{array}{c}
\frac{\delta H_{n+1}}{\delta
u}\\
\frac{\delta H_{n+1}}{\delta
v}\\
 \end{array}\right)=K\left(\begin{array}{c}
\frac{\delta H_{n}}{\delta
u}\\
\frac{\delta H_{n}}{\delta
 v}\\
 \end{array}\right)$$
where $$J=\left(\begin{array}{cc}
0 & \partial\\
\partial &0\\
 \end{array}\right),~K=\left(\begin{array}{cc}2\partial & \partial^{2}+\partial u\\
-\partial^{2}+u\partial  & \partial v+v \partial\\
 \end{array}\right),$$$$\left(\begin{array}{c}
a_{n+1}\\
c_{n+1}\\
 \end{array}\right)=L\left(\begin{array}{c}
a_{n}\\
c_{n}\\
 \end{array}\right),~L=\left(\begin{array}{cc}
\partial^{-1}u\partial-\partial & \partial^{-1}v\partial+v\\
2&\partial+u\\
 \end{array}\right),~n=1,2,\cdots$$
 $$a_0=\frac{1}{2},~b_0=c_0=0,~a_1=0~,b_1=v,~c_1=-1,\ldots.$$
 For $N$ distinct real $\lambda_{j}$, from the spectral problem
\begin{equation}
\label{ckv2} \left(\begin{array}{c}
\psi_{1j}\\
\psi_{2j}\\
 \end{array}\right)_{x}=\left(\begin{array}{cc}
-\frac{1}{2}(\lambda_{j}-u) & -v\\
1 & \frac{1}{2}(\lambda_{j}-u)\\
 \end{array}\right)\left(\begin{array}{c}
\psi_{1j}\\
\psi_{2j}\\
 \end{array}\right),
\end{equation}
we have
\begin{equation}
\label{lamdac}\frac{\delta \lambda_{j}}{\delta
u}=\psi_{1j}\psi_{2j},~ \frac{\delta \lambda_{j}}{\delta
v}=-\psi_{2j}^{2}.\end{equation}
When take
$\gamma_{j}=-\lambda_{j},~\mu_{j}=-1,$
 and $n=3$ in (8),  the
  new generalized coupled KdV equation is
constructed  as follows
\begin{subequations}\label{gckdv} \begin{align}
&u_{t}=u_{xxx}+6u_{x}v+6uv_{x}+3u_{x}^{2}+3uu_{xx}+3u^{2}u_{x}-\sum\limits_{j=1}^{N}\omega_{2jx},\\
&v_{t}=v_{xxx}+6uu_{x}v+6vv_{x}+3u^{2}v_{x}-3uv_{xx}-\sum\limits_{j=1}^{N}\omega_{1jx},\\
&2\omega_{1jx}+\omega_{2jxx}+u_{x}\omega_{2j}+u\omega_{2jx}-\lambda_{j}\omega_{2jx}=0,\\
&-\omega_{1jxx}+u\omega_{1jx}+v_{x}\omega_{2j}+2v\omega_{2jx}-\lambda_{j}\omega_{1jx}=0,~j=1,2,\cdots,N.
\end{align}
\end{subequations}
The generalized Kupershmidt deformation of the cKdVH is constructed
as follows
\begin{subequations}
\label{eqns:ckdv3}
\begin{align}
&\left(\begin{array}{c}
u\\
v\\
 \end{array}\right)_{t_{n}}=J(\left(\begin{array}{c}
\frac{\delta H_{n+1}}{\delta
u}\\
\frac{\delta H_{n+1}}{\delta
 v}\\
 \end{array}\right)
 -\sum_{j=1}^{N}\left(\begin{array}{c}
\frac{\delta \lambda_{j}}{\delta
u}\\
\frac{\delta \lambda_{j}}{\delta
v}\\
 \end{array}\right))
 ,\label{eqns:ckdv3a}\\
& (K-\lambda_{j}J)\left(\begin{array}{c} \frac{\delta
\lambda_{j}}{\delta
u}\\
\frac{\delta \lambda_{j}}{\delta
v}\\
 \end{array}\right)=0,~j=1,2,\cdots,N.\label{eqns:ckdv3b}
\end{align}\end{subequations}
From (\ref{lamdac}) and (\ref{eqns:ckdv3b}), we obtain
\begin{subequations}
\label{eqns:ckdv4}
 \begin{align}
 \nonumber&2\psi_{1jx}\psi_{2j}+2\psi_{1j}\psi_{2jx}-2\psi_{2jx}^{2}-2\psi_{2j}\psi_{2jxx}-u_{x}\psi_{2j}^{2}\\
 &-2u\psi_{2j}\psi_{2jx}
 +2\lambda_{j}\psi_{2j}\psi_{2jx}=0,\label{eqns:ckdv4a}\\ \nonumber
&
-\psi_{1jxx}\psi_{2j}-2\psi_{1jx}\psi_{2jx}-\psi_{1j}\psi_{2jxx}+u(\psi_{1jx}\psi_{2j}+\psi_{1j}\psi_{2jx})-v_{x}\psi_{2j}^{2}\\
&-4v\psi_{2j}\psi_{2jx}
 -\lambda_{j}(\psi_{1j}\psi_{2jx}+\psi_{1jx}\psi_{2j})=0. \label{eqns:ckdv4b}
 \end{align}
\end{subequations}
(\ref{eqns:ckdv4a}) yields
\begin{equation}
\label{eqns:36}
\psi_{2jx}=\psi_{1j}+\frac{1}{2}(\lambda_{j}-u)\psi_{2j}+\frac{\mu_{j}}{\psi_{2j}},
~j=1,2,\cdots,N.\end{equation}
(\ref{eqns:ckdv4b}) yields
$$
[\psi_{2j}(\psi_{1jx}+\frac{1}{2}(\lambda_{j}-u)\psi_{1j})+v\psi_{2j}]_{x}+(\frac{\psi_{1j}\mu_{j}}{\psi_{2j}})_{x}+2\psi_{1jx}(\psi_{2jx}
-\frac{\mu_{j}}{\psi_{2j}}) $$
\begin{equation}
\label{eqns:37}
+\psi_{2jx}(2v\psi_{2j}+u\psi_{1j}+\lambda_{j}\psi_{1j})=0.
\end{equation}
In order to keep (\ref{eqns:37}) hold, we find that $\mu_{j}
(j=1,\cdots,N)$ equal zero.  So (\ref{eqns:ckdv3b}) gives
$$\psi_{1jx}=-\frac{1}{2}(\lambda_{j}-u)\psi_{1j}-v\psi_{2j},~\psi_{2jx}=\psi_{1j}+\frac{1}{2}(\lambda_{j}-u)\psi_{2j},~j=1,2,\cdots,N.$$
Then the generalized Kupershmidt deformation of coupled KdV equation
gives rise to the following  system
\begin{subequations}\label{ckdv5} \begin{align}
&u_{t}=u_{xxx}+6u_{x}v+6uv_{x}+3u_{x}^{2}+3uu_{xx}+3u^{2}u_{x}+2\sum\limits_{j=1}^{N}\psi_{2j}\psi_{2j,x},\\
&v_{t}=v_{xxx}+6uu_{x}v+6vv_{x}+3u^{2}v_{x}-3uv_{xx}-\sum\limits_{j=1}^{N}(\psi_{1j}\psi_{2j})_{x},\\
&\psi_{1j,x}=-\frac{1}{2}(\lambda_{j}-u)\psi_{1j}-v\psi_{2j},~\psi_{2j,x}=\psi_{1j}+\frac{1}{2}(\lambda_{j}-u)\psi_{2j},~j=1,2,\cdots,N,
\end{align}
\end{subequations}
which is  called as the coupled KdV equation with self-consistent
sources.  Following the procedure in \cite{Roso5,Roso6,cb2},
 we can find  the Lax representation (\ref{hd1}) and
(\ref{eqns:v}) for (\ref{ckdv5}) with
$$U=\left(\begin{array}{cc}
-\frac{1}{2}(\lambda-u) & -v\\
1 & \frac{1}{2}(\lambda-u)\\
 \end{array}\right),$$
$$V=\left(\begin{array}{cc}
 A&B\\C&-A\\
 \end{array}\right)
 +\frac{1}{2}\sum\limits_{j=1}^{N}\left(\begin{array}{cc}
 \psi_{2j}^{2}& 0\\
0 &-\psi_{2j}^{2}\\
 \end{array}\right)-\sum\limits_{j=1}^{N}\frac{1}{\lambda-\lambda_{j}}
 \left(\begin{array}{cc}
 \phi_{1j} \phi_{2j}& - \phi_{1j}^{2}\\
 \phi_{2j}^{2}&-\phi_{1j} \phi_{2j}\\
 \end{array}\right)$$
 where
 $$A=\frac{1}{2}\lambda^{3} +v\lambda-4uv-v_{x}-u_{xx}-3uu_{x}-u^{3}$$
 $$B= v\lambda^{2}+(uv-v_{x})\lambda+2v^{2}+u^{2}v-2uv_{x}-u_{x}v+v_{xx}$$ $$C=-\lambda^{2}
-u\lambda-u_{x}-2v-u^{2}.$$

\section{Conclusion}
 The main purpose of this paper is to show that for the bi-Hamiltonian systems with both Hamiltonian operators being differential operators,
 the generalized Kupershmidt deformation (GKD) developed from the Kupershmidt deformation in \cite{kd} offers an useful way to construct
 new integrable systems starting from
 bi-Hamiltonian systems. We construct some new integrable systems by
making use of  the generalized Kupershmidt deformation (GKD) of
bi-Hamiltonian systems and to verify the conjecture on the
integrability of the generalized Kupershmidt deformation in some
specific cases.  We obtain the new GKD of Harry Dym equation, GKD of
the classical Boussinesq equation and GKD of the coupled KdV
equation. Then we show that these new systems  can be converted into
the Rosochatius deformation of soliton equation with self-consistent
sources. Furthermore the Lax pairs for the Rosochatius deformation
of soliton equation with self-consistent sources can be constructed
in a systematic procedure.  These  imply that the generalized
Kupershmidt deformation of bi-Hamiltonian systems provides a new way
to construct new integrable systems from bi-Hamiltonian systems and
also offers a new approach to obtain the Rosochatius deformation of
soliton equation with self-consistent sources. However, when the
Hamiltonian operators are not pure differential operators, it
remains to study how to construct new integrable
  systems from bi-Hamiltonian systems by using the generalized
  Kupershmidt deformation.

\section*{Acknowledgement}
This work is supported by National Basic Research Program of China
(973 Program) (2007CB814800), National Natural Science Foundation of
China (10901090,10801083) and  Chinese Universities Scientific Fund
(2011JS041).

\end{document}